\documentclass[a4paper,twocolumn,12pt]{article}

\usepackage[top=25mm,bottom=25mm,left=25mm,right=25mm]{geometry}

\usepackage{amsmath} 
\usepackage{graphicx}
\usepackage{epstopdf}
\usepackage{pdfpages}
\usepackage{url}
\usepackage{setspace} 
\usepackage{float}
\usepackage{gensymb}
\usepackage{titlesec}
\titleformat{\section}{\bfseries\large\scshape\filcenter}{\thesection}{1em}{}
\titleformat{\subsection}{\bfseries\normalsize\scshape\filcenter}{\thesubsection}{1em}{}
\usepackage{gensymb}
\usepackage[english]{isodate}

\newcommand{\captionfonts}{\footnotesize}
\renewcommand\thesection{\Roman{section}.}
\renewcommand\thesubsection{\Alph{subsection}.}

\makeatletter
\long\def\@makecaption#1#2{
  \vskip\abovecaptionskip
  \sbox\@tempboxa{{\captionfonts #1: #2}}%
  \ifdim \wd\@tempboxa >\hsize
    {\captionfonts #1: #2\par}
  \else
    \hbox to\hsize{\hfil\box\@tempboxa\hfil}%
  \fi
  \vskip\belowcaptionskip}

\renewcommand\p@subsection{\thesection}
    
\makeatother

\begin{document}
\setlength{\parskip}{0pt}
\title{Precursor Photometric Study of Astras Satellite Cluster 19.2$\degree$E for Geostationary Debris Profiling}

\author{Toyaj Singh}
\date{\small\printdayoff June 2019}
\vspace{-10mm}

\twocolumn[
\maketitle 
\vspace{-10mm}
\begin{@twocolumnfalse}
\begin{abstract} 
\noindent
Aperture photometry was performed on images (from the SuperWASP instrument) of the 19.2$\degree$E geostationary Astras satellite constellation over 5 nights to form light curves and predict model and movement parameters of the satellites. The 1KR satellite is observed to have two peaks, 90$\degree$ out of phase indicating rotating wings, while satellite 1M is predicted to have shadowing due to dips in its light curve or a single wing design due to a single major peak. All satellites have peak magnitudes at the minimum sun phase angle 28$\degree$ as expected. The developed methods and results serve a precursor to developing methods to address the ultimate goal of geostationary debris profiling.    
\end{abstract}
\vspace{11mm}
\end{@twocolumnfalse} ]

\maketitle

\section{Introduction}
Satellites are perhaps now more commonplace and essential than they have ever been. With the expansion into the information age[1], communications and technology have become increasingly fundamental to day to day activities, stemming from recreation to research. Naturally this causes a dependency on satellites and it is logical to consider the implications of the failure or malfunction of these quintessential bodies.

Satellites that reside on the geosynchronous Earth orbit (GEO) are of particular significance to communications for military[2] and weather forecasting[3]. Geosynchronous satellites maintain the same period with Earth resulting in them always returning to the same position over the sky each sidereal day (complete rotation with respect to fixed stars). Geostationary satellites are a particular type of geosynchronous satellite that are positioned above the equator resulting in them remaining fixed above a designated region relative to the Earth [4]. This unique property is what makes these satellites so invaluable in the aforementioned communications area, as transmitter and receivers do not need to track them. But understandably the limited size of the geostationary belt makes it extremely desirable and means setting up these geostationary satellites is expensive [5]. Due to this financial and societal interest, the security of these satellites is of worthwhile scrutiny.

\begin{figure}[t]
\centering
\includegraphics[width=77mm]{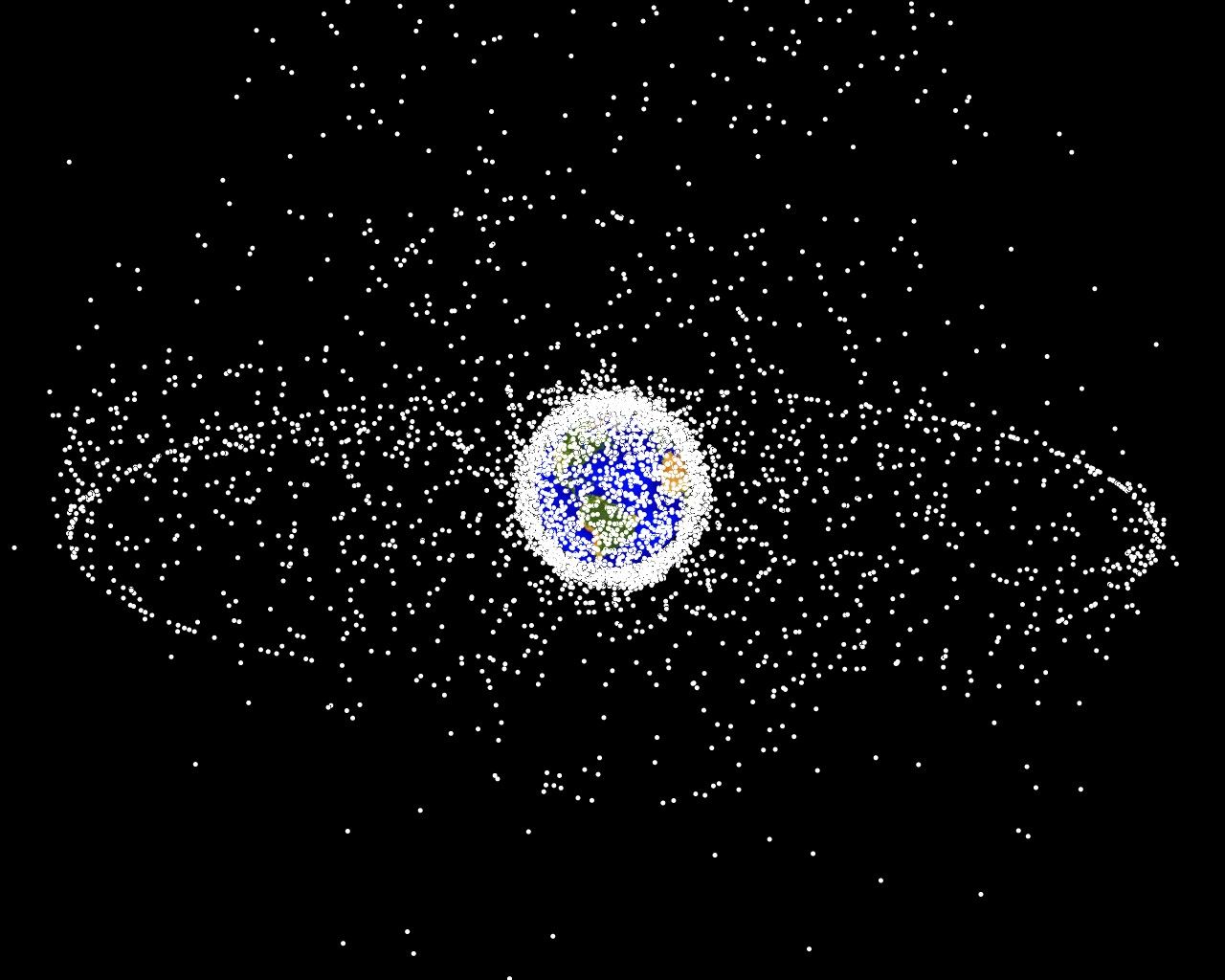}
\vspace{-6mm}
\caption{\small A model of satellite and debris population of the GEO belt. Each dot represents an object. The higher density ring is of the geostationary belt and the higher population of objects in the Northern hemisphere is due to high inclination orbit Russian objects. Note the dots have been enlarged and are not to scale with Earth. Image courtesy of NASA.}
\label{fig: fig_1}
\end{figure}

Geostationary satellites have a typical lifespan of 7 years [6] which is variant on failure and operational power. They are dependent on manoeuvring in order to keep their position and orientation in the desired formation. These manoeuvres expend fuel which puts a limitation on the lifespan of these satellites, after which they are typically removed from the geostationary belt. However, this natural cycle assumes non-failure.  The 2 main methods of failure for geostationary satellites are classified as gradual and catastrophic. Gradual degradation is imminent from the launch of the satellite due to solar effects and is difficult to address after the satellites’ launch. Catastrophic failure is of more concern, especially recently with the rise in resident space objects(RSO)[7], leading to a higher estimated risk of collision between debris and operational geostationary satellites. As of late 2018 there are 558 active GEO satellites in orbit, as per the UCS Satellite Database [8], which indicates the density of the belt without even considering the more substantial small ($<$5 cm) debris population. Fig \ref{fig: fig_1} shows a model of the debris population around the GEO belt of Earth as modelled by NASA’s Orbital Debris Program [9]. The origins of this debris range from “slag and dust from solid rocket motors” and “surface degradation products”[10] such as paint fragments to coolants and needle clusters for creating an artificial ionosphere[11][12]. But a large proportion of this debris is simply from old nonoperative geostationary satellites. 

It is evident from the density of the debris population why collisions are a non-trivial probability, and from the 2009 collision of Kosmos-2251 and Strela[13] the debris population aggregated even further, with NASA estimating approximately 1000 fragments of debris larger than 10 centimetres being expelled. The danger of even minuscule debris is considerable due to the high velocities achieved. Despite this the real danger with debris is not its population but the limited accuracy with which it can be identified. The uncertainty of the debris population to a microscopic degree is what drives the motivation for this research.

\begin{figure*}[t]
\centering
\includegraphics[width=77mm,height=50mm]{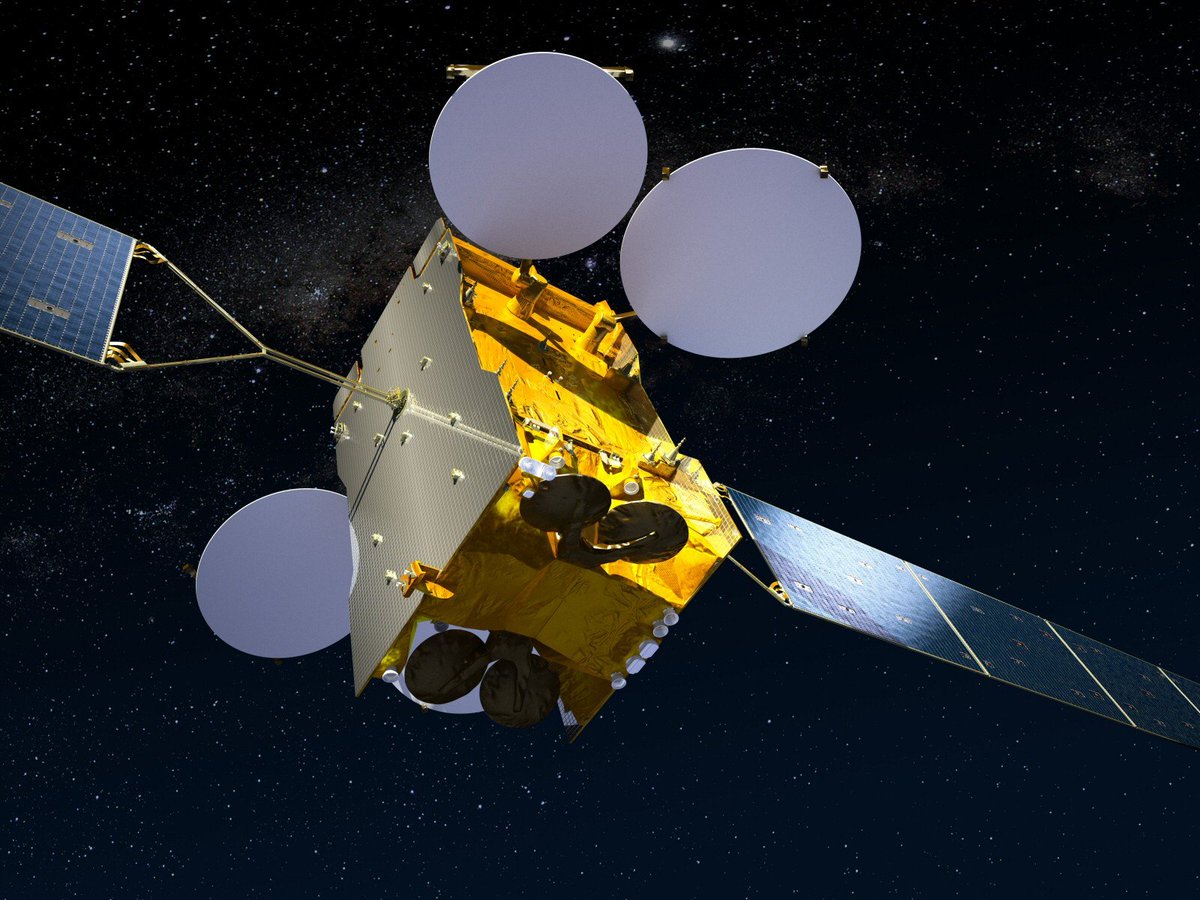} \hspace{3mm}
\includegraphics[width=77mm,height=50mm]{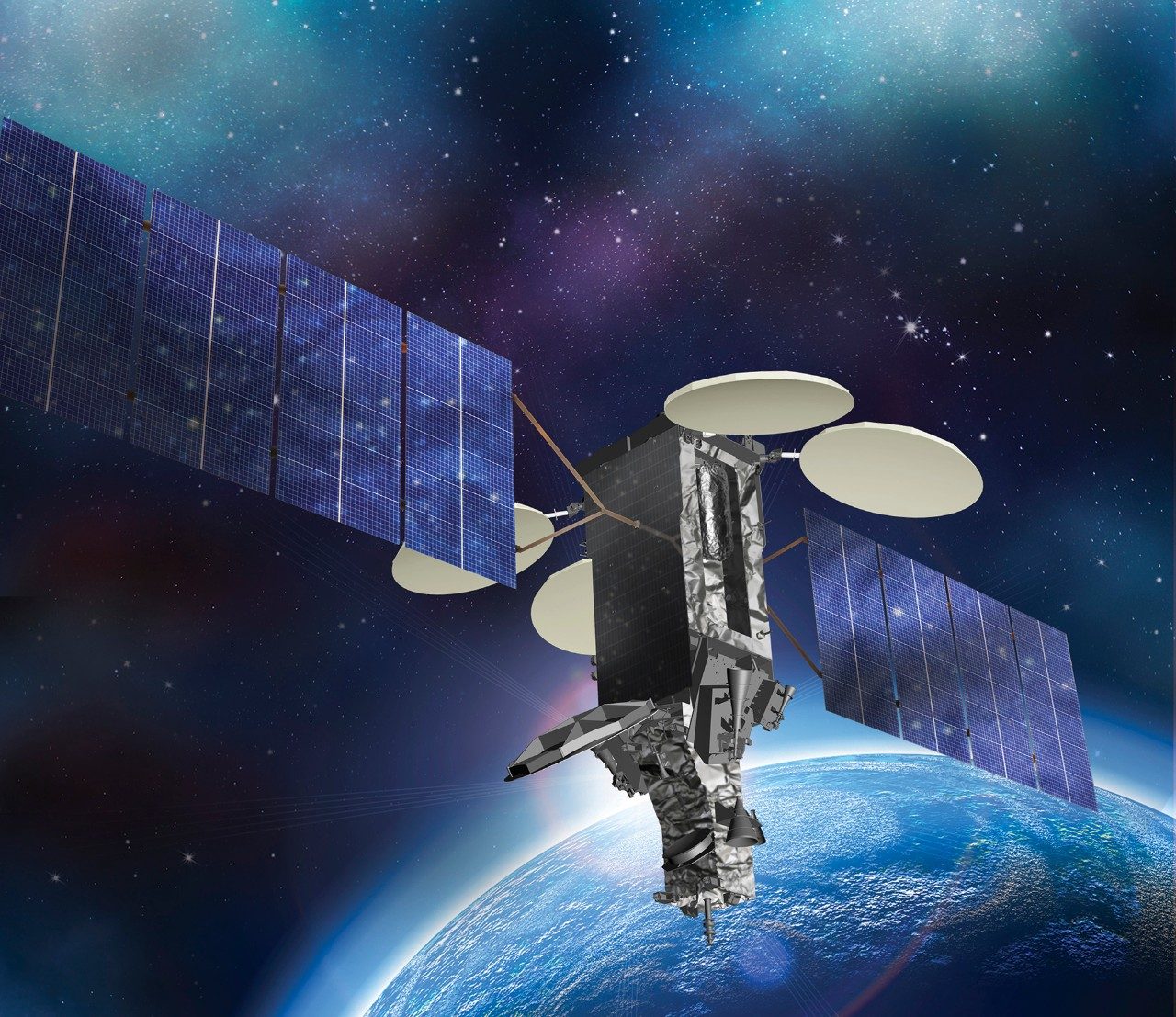}
\vspace{-5mm}
\caption{\small 2a on the left is a digital model of Eurostar's elementary E3000 model. 2b on the right is a digital model of Lockheed Martin's elementary A2100 model. Models courtesy of Eurostar[15] and Lockheed Martin respectively[16] }
\label{fig: fig_2}
\end{figure*}

The aim of this study is to observe a set of communication geostationary satellites known as the Astras[14] and develop methods to study their behaviour based off their movement and flux variations and then utilise the acquired methods to study debris profiles in the geostationary belt.  The Astras satellites are geostationary communication satellites operated by SES S.A., a global video and data broadcasting firm. The Astras consist of 4 satellites (1M, 1L, 1KR, 1N) built from 2 models each: Airbus's Eurostar E3000 and Lockheed Martin's A2100 as shown in Fig \ref{fig: fig_2}. Due to the commercial confidentiality associated with the satellites, it was not possible to have prescience of the exact design and dimensions of the satellites which ultimately made it difficult to accurately explain the features observed in the light curves. The schematics in Fig \ref{fig: fig_2}[15][16] gave a rough visualisation of the general models however the specific modifications and dimensions remained unknown. 

This project will be working with Rowe-Ackermann Schmidt Astrograph(RASA)[17] data, located in La Palma as well as data from  SuperWASP(Wide Angle Search for Planets)[18] also in La Palma. The aim is to generate flux profiles over time for each of the 4 Astras, formally known as `light curves' and identify each satellite using known two line element(TLE) information. From these `light curves' the goal is to identify features of significance and use them to make conclusions about the design and behaviour of the satellites as well as develop methodology that can be transferred for debris observation with appropriate data.

The condition of the geostationary belt is undoubtedly in a precarious position, and this project aims to study the behaviour of geostationary satellites in detail to see the effects of debris on observable satellite properties in the hopes to understand the significance and cause of their behaviour. Eventually the goal would be to replicate these methods with appropriate data of adequate optical precision, such as that from the Isaac Newton Telescope[19], to profile the debris in the geostationary belt and accurately determine position profiles at present and future times.	

\section{Background}
\subsection{Geostationary Satellites}
An ideal geostationary satellite would remain fixed above the equator with an orbit of inclination 0$\degree$ and a period of 1436\textsuperscript{m}.07 [20] which can be calculated by Kepler's Third Law given by \eqref{eq: Eq_1}:

\begin{equation}
T^{2}=\frac{4\pi^{2}}{GM}a^{3}
\label{eq: Eq_1}
\end{equation}
where T represents the sidereal period[21,22] of the orbiting body, G the universal gravitational constant, M the mass of the Earth and a the semi-minor axis of the Earth.

However in practice, geostationary satellites move in small orbits of inclination which makes them appear to be oscillating in declination across the sky[20]. Additionally, due to the non-uniformity of the Earth's gravitational field as well as the effects of the moon and sun's gravitation fields, geostationary satellites are observed to drift down from their declination. As mentioned, operational satellites often compensate for this through propulsion [23] however this is limited by the fuel available. 

It is typical for retired geostationary satellites to be removed from the GEO belt using their remaining fuel, however this practice has only recently become ubiquitous. Retired satellites that remain in the GEO belt, referred to as `librational'[24], follow an elliptical pattern around the nearest stable point and their inclination varies from 0\degree to 15\degree [25]. This corresponds to these librational satellites crossing the equatorial plane twice a year with velocities peeking at 800 m/s[24]. Naturally this shows the risk of collision from retired satellites alone, let alone smaller debris, which despite being much lighter, can reach significantly higher velocities[26].

\subsection{CCDs}
\begin{figure}[b]
\centering
\includegraphics[width=77mm]{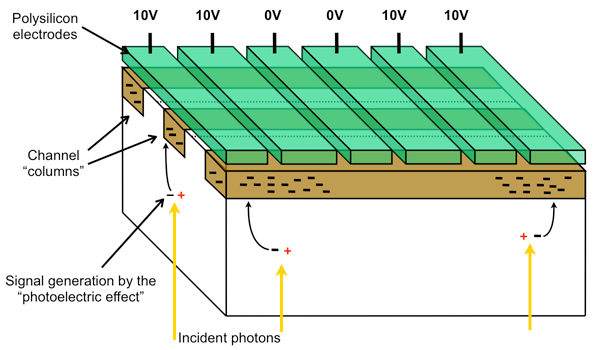}
\vspace{-6mm}
\caption{\small A schematic of a CCD chip cross section. The lattice of electrodes and the adjacent channel columns are shown as well incident photons and the resulting photoelectrons. Courtesy of Gaia UK[31] }
\label{fig: fig_3}
\end{figure}

The main method of detecting and recording images of faint objects such as GEO satellites is through the use of charge coupled devices (CCDs). CCDs are integrated circuits built on a silicon surface consisting of pixel arrays that form potential wells from “applied clock signals” to store and deliver packets of charge[27]. Typically, the charge packets are composed of electrons resulting as a consequence of the photoelectric effect[28] from incident photons on the silicon body or the apparatus’ internal dark signal[29]. CCDs accumulate electronic charge at each cell (pixel) over the duration of exposure (integration time) and then transport the charge through the chip where it is read from an “analogue-to digital” converter(ADC)[30] to give a binary measurement. The charge packets are moved through the use of a time-variable voltage which shifts the charge to the ADC where it can be converted into a voltage reading.  Fig \ref{fig: fig_3} demonstrates how a CCD is constructed as well how the the signal is generated from the photoelectrons being emitted.

Due to the faint flux profiles of astronomical objects, especially in the relevant case of debris, it is necessary to have detectors with a diminutive waste of photons. CCDs achieve this immaculately with quantum efficiencies over 90\% [32]. The quantum efficiency is the ratio of incident photons to the converted electrons as given by \eqref{eq: Eq_2} where n\textsubscript{e} is the number of electrons collected (per second) and n\textsubscript{p} is the number of incident photons (per second).

\begin{equation}
QE=\frac{n_e}{n_p}
\label{eq: Eq_2}
\end{equation} 

This is why CCDs supplanted their predecessor: photographic film which could only achieve quantum efficiencies of around 10\%.

Due to the reliance of CCDs on the photoelectric effect, there is a limit on the the minimum wavelength they can detect. Electron emission is only viable when the incident photons have enough energy to move the electrons from the valence band to the conduction band in the silicon CCD [33]. From Planck's relation shown in \eqref{eq: Eq_3} (where E is the energy of the incident photon, h is Planck's constant and f the frequency of the incident light) it can be seen there is a natural threshold wavelength for detection by CCDs. 

\begin{equation}
E=hf
\label{eq: Eq_3}
\end{equation}

The threshold is typically $~1100$ nm for silicon based CCDs[34].  

The use of  CCDs is under some scrutiny with the rise of Active Pixel Sensor (APS) technology [35] due to CCD's problems with ionisation and displacement damage[36]. However, they still remain omnipresent in modern detectors and this is the case for RASA and SuperWASP and hence this project. 

\subsection{RASA, SuperWASP and Corrections}

\begin{figure}[t]
\centering
\includegraphics[width=77mm]{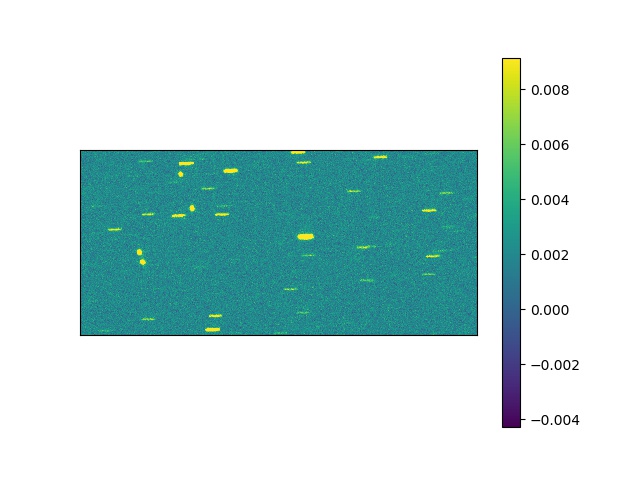}
\vspace{-6mm}
\caption{\small A raw uncorrected image taken from the RASA telescope mapped to a logarithmic colour-scale. The trails are stars due to the fixed telescope, whilst the 4 dots are the geostationary satellites (Astras) }
\label{fig: fig_4}
\end{figure}

The primary source of data for this project comes from the northern telescope array of SuperWASP (Wide angle search for planets) located in La Palma, Canary Islands; although for preliminary work, data from the RASA telescope will be used. The images taken from RASA and SuperWASP are taken using the aforementioned CCDs and have an observable background profile. Correcting this background is essential for any further methods in extracting information from the images. Fig \ref{fig: fig_4} shows a raw CCD image from the RASA telescope mapped to a logarithmic colour-scale with the 4 Astras satellites visible as dots. CCD images require 3 main corrections bias, dark current and flat field.

Due to the inherent noise associated with CCD imaging it is possible to observe `negative' counts. To compensate for this, an exposure (or multiple) with zero integration time is taken so that no photoelectrons are recorded. This bias frame can then be subtracted from the real image. Often it is beneficial to take multiple bias frames and calculate an absolute mean bias frame and subtract that from all images. 

As mentioned before, CCD imaging is enabled by the emission of photoelectrons moving from the valence band to the conduction band. However in detection, the readings are corrupted by thermal electrons, which is why CCDs are typically cryogenically cooled. Despite this, it is normal for images to still be affected by this so called `dark current'. To mitigate this effect, exposures with a closed shutter are taken so that only the dark current is detected. This gives a dark frame which can then be subtracted from all the data images. As the dark frame also captures the natural noise in detection, it must also have the bias frame subtracted before being deducted from the image data. For the purpose of this project, dark current is assumed to be negligible due to the exposure time used and cooling of the RASA and SuperWASP instruments; hence no dark current subtraction is performed.

The final correction: flat field, corrects for three significant effects: vignetting[37], pixel to pixel variations[38] and dust on optics. Vignetting is the natural light loss near the periphery of an image compared to its centre due to mechanical and optical effects within the apparatus. Pixel to pixel variations are caused by the natural variance in the size of the individual pixels in the CCD, which leads to different photon counts for the same exposure. Finally, dust on optics simply refers to dust particles on the CCD window, resulting in dark `donuts' being cast on the image data. To enable flat field corrections, an image is taken of the twilight sky, which is assumed to be uniform so that only the aforementioned factors lead to variation. The image must be taken using the filter with which the data was collected in order to compensate for the dust effect. The data images are then divided by the flat field image, after it has also been bias reduced. 

This project will be utilising bias correction methods as well as flat field ones. Additionally, background subtraction[39] will be utilised but only through a relevant computational model rather than from immediate open sky exposures at the time of study, which is the typical method.  

\subsection{Astrometry}
To observe the Astras satellites (and subsequently debris) of interest, it is essential to determine their absolute positions, defined by the right ascension (RA) and declination(Dec). This is achieved through astrometry using the NASA-endorsed software: `astrometry.net'[40]. 

The software takes input image data and calibrates the field of view by comparing the relevant sources in the image to a known catalogue of index stars. The software returns a calibrated header, the x,y position data for all the sources in the image and the desired transformation from x,y to RA and DEC. The software can also take as input a tabular array of the x and y positions of the sources and use them to calibrate a correct translation into RA and Dec. For this project the latter method will be utilised due to its improved computational efficiency and accuracy. This improved accuracy is justified by the ability to set parameters on what is classified as the source stars, as they appear as trails in the fixed images rather than points. 

\subsection{Photometry}
In order to study the behaviour of the Astras satellites, it is necessary to observe their observable nature over time. Naturally the flux variations are an appropriate property to visualise and study. The flux profile of the relevant sources (Astras) can be plotted over the study time leading to light curves that detail how the flux varies throughout the day for each satellite. There are two central photometry methods employed in modern research: aperture photometry[41] and point spread function photometry[42]. For aperture photometry, an aperture must be placed over each object and the total pixel counts within the aperture are summed. Then the product of the average background pixel count and number of pixels in the aperture are deducted from the sum. Point spread photometry on the other hand relies on representing every source in the image by a point spread function. If the ``function is not spatially invariant it can be assumed that it is knowable and can be modelled"[43] and this allows the flux values for each source to be determined. Point spread photometry offers higher precision especially for low magnitude objects where aperture photometry proves inadequate. However, it is often more computationally expensive and more complex to implement. For this project aperture photometry is adequate due to geostationary satellites being relatively bright in retrospect to distant interstellar objects, allowing an acceptable degree of precision. The resulting flux from either of these photometric methods can then be used to plot light curves for each object (satellite). This project utilises this in order to directly compare the similarities and differences in the behaviours of the satellites, which are expected to behave similarly due to their similar design structure[10] and positional similarities. 

\subsection{Light Curves}
Light curves profile the brightness of an object over some period of observation. Typically photometry allows the flux evolution to be recorded which can then be calibrated and converted into a visual magnitude scale and plotted to give a continuous light curve. The use of light curves for astronomical study has been commonplace due to their reliability and reproducibility. By knowledge of the light curves of a standard library of objects, it is possible to categorise and predict the features of unknown objects which is what this project aims to do at a preliminary phase. The eventual aim to address the debris profiling problem will be to form a standard catalogue for geostationary satellites in regular operation as well as post collision with debris. Typically light curves have been useful in studying exoplanets and binary star systems however recent research has been conducted in utilising them for satellites and in particular geostationary[44] ones. Modern developments in light curve inversion have furthered the use of light curves, however photometry has been sufficient for shape and spin determination for most academic purposes. The benefits of light curve inversion come into fruition for ``pole orientation, rotation period, three-dimensional shape, and scattering properties of the surface of a body"[45] as addressed in J. Torppa's 2001 paper on asteroid light curves. Although this technique since its inception in 2001 has only been used for light curves of asteroids and more distant natural bodies, recently it has been utilised on GEO objects[46] using ``space based sensors" as detailed in B. Bradley's study on `light curve inversion for shape estimation of GEO objects'. However, Bradley's method relied on prescience of the objects (satellites) albedos' and materials, which for this project is not possible. 

\vspace{-2mm}
\subsection{Two Line Elements}
Orbital information for Earth orbiting objects are typically encoded in a ``two line element format"(TLE)  which contain the parameters associated with the satellite in two rows with many columns. Two line elements are used with ``Simplified General Perturbations models"(SGP4)[47] to predict the position and velocity of orbital objects. These models predict the effect of perturbations by the Earth's gravitational field as well as that of the sun and moon on orbital objects. The error associated with the SGP4 method is 1 km at epoch (most recent time of measurement) and grows 2-3 km per day. TLEs are updated frequently and so this error is rarely too significant for accurate consideration. TLEs were useful in this experiment to identify which light curve corresponded to which satellite, which in turn was useful to understand the behaviour of the light curves and the effect of model type on them. 

\section{Methodology}
The immediate aim of the project was to produce light curves of the Astras satellite cluster. To achieve this, data from the SuperWASP (and RASA initially) was corrected for bias and flat fielding effects through aggregate bias and flat frames. From the corrected frames it was possible to perform the aforementioned astrometry and photometry through python methodologies assisted by the astropy[48] and sep[49] modules. This produced positional plots of the satellites as well as the main light curves.

Additionally, it was necessary to identify the Astras satellites individually which was achieved through accessing TLEs of the appropriate epoch for the 4 satellites and creating predicted positional plots in astronomical coordinates. These could be compared directly to the positional plots from the SuperWASP data and allowed determination of the satellites.

The later goals of the project were to develop methods to study debris however, due to time scarcity and gratuitous results from the light curves, the primary goal shifted to trying to explain the features of the light curves and make predictions about the Astras satellite models. 

\begin{figure}[t]
\centering
\includegraphics[width=77mm]{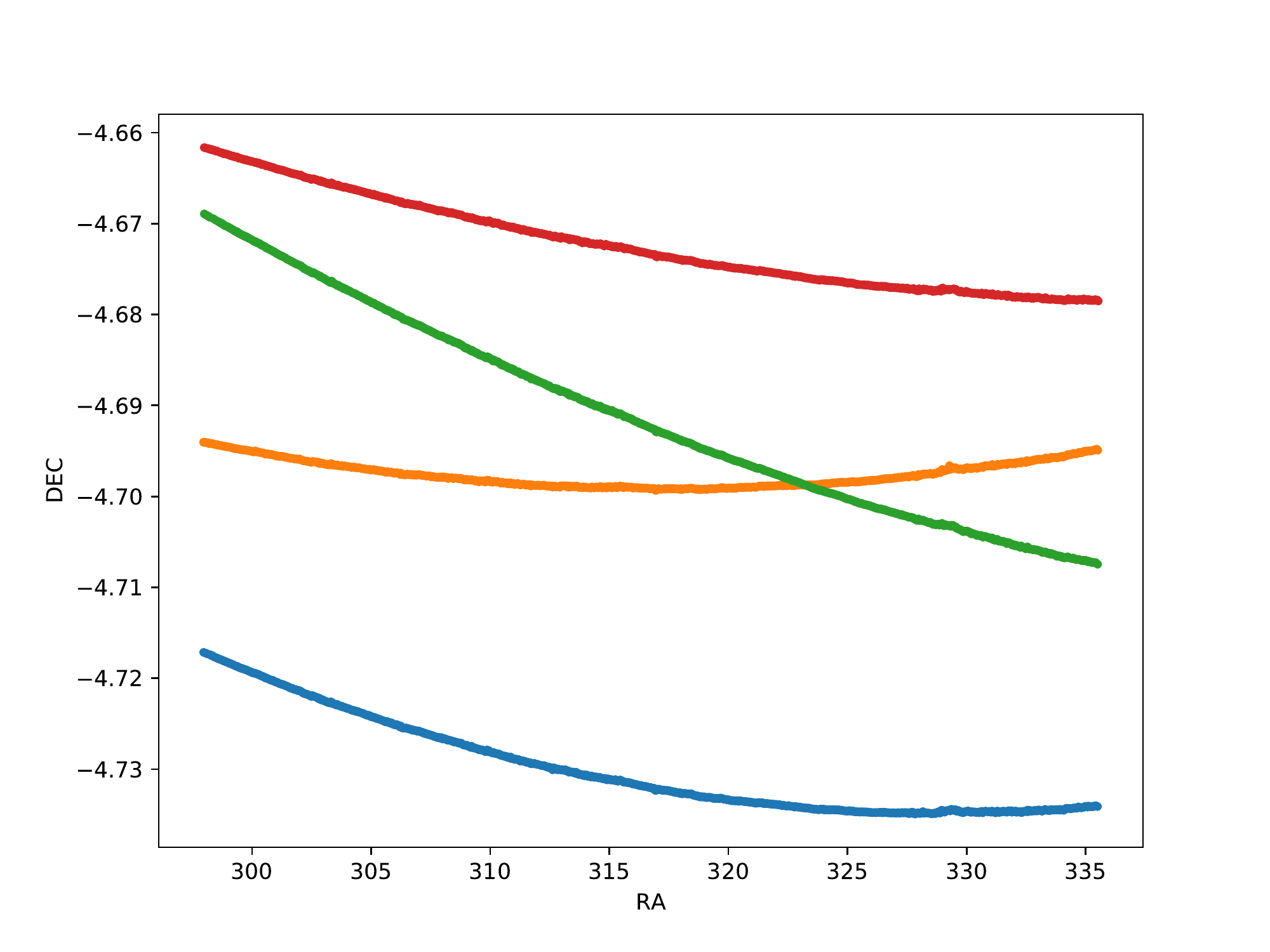}
\vspace{-6mm}
\caption{\small Plot of declination against right ascension for the 4 Astras satellites over a Summer night. Observed by the RASA telescope in La Palma }
\label{fig: fig_5}
\end{figure}

\section{Results and Discussion}
\subsection{RASA Light Curves}
The RASA instrument was used to collect data over a 3.5 hour period on the 21st of June 2018 of the Astras cluster and were saved in the form of FITS files. These images were corrected for the aforementioned bias and flat field effects and through astrometry (via astrometry.net) were used to produce a map for the progression of the satellites throughout the period of observation in astronomical coordinates as shown in Fig \ref{fig: fig_5}. Due to the smaller angular scale of pixels of the RASA the positional plots were constructed with negligible blending as observed by the continuous plots in Fig \ref{fig: fig_5}.

\begin{figure}[t]
\centering
\includegraphics[width=77mm]{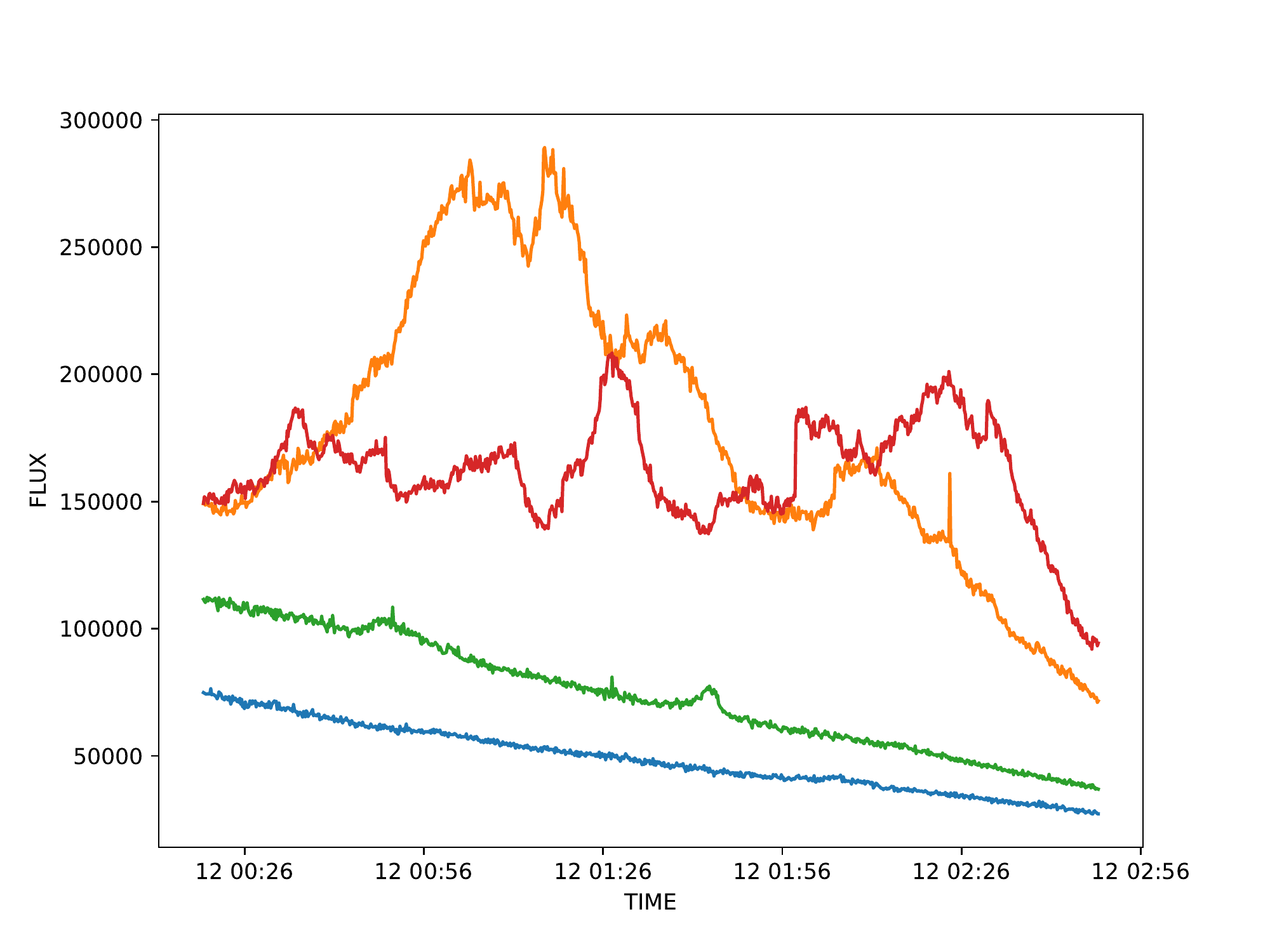}
\vspace{-6mm}
\caption{\small Light curves for the 4 Astras satellites over a Summer night. Observed by the RASA telescope in La Palma }
\label{fig: fig_6}
\end{figure}

Further, from the image set (through a python enabled aperture photometry kit) the flux counts over time were extracted. From this, light curves were constructed as shown in Fig \ref{fig: fig_6} where each coloured line represented a separate Astras satellite. The results are immediately surprising due to the 2 distinct polarising trends observed by the light curves. The natural descent of the absolute magnitude for all 4 satellites is expected through the night, however the erratic behaviour of 2 of the satellites (1L and 1KR) was of particular interest.   

\begin{figure*}[]
\centering
\includegraphics[width=17cm, height=23cm]{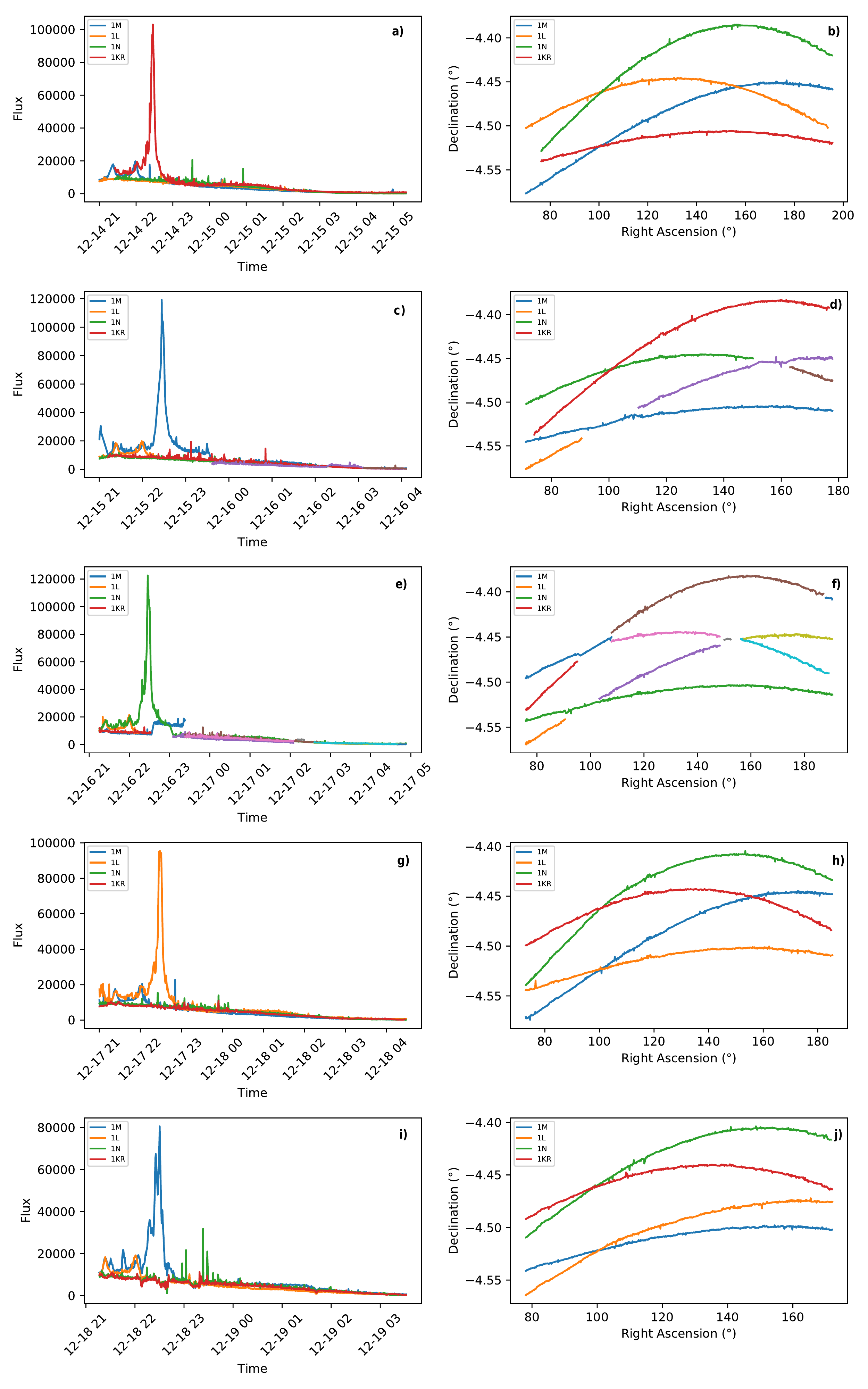} \hspace{3mm}
\vspace{-6mm}
\caption{\small Plots a, c, e, g, i show light curves for the 4 Astras satellites over 5 consecutive Winter nights. Plots b, d, f, h, j show declination against right ascension for the 4 Astras satellites over 5 consecutive Winter nights. Each colour corresponds to the same satellite excusing blending effects. Observed by the SuperWASP instrument in La Palma }
\label{fig: fig_7}
\end{figure*}

\begin{figure}[t]
\centering
\includegraphics[width=80mm]{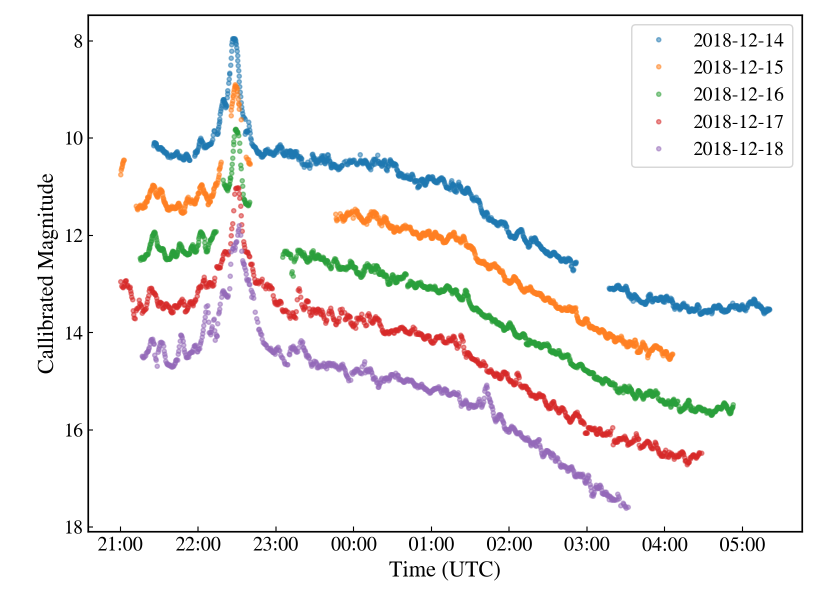}
\vspace{-6mm}
\caption{\small Calibrated light curves for Astra satellite 1KR for 5 consecutive Winter nights. Observed by the SuperWASP instrument in La Palma }
\label{fig: fig_8}
\end{figure}

\subsection{SuperWASP Light Curves}
\begin{figure}[!b]
\centering
\includegraphics[width=80mm]{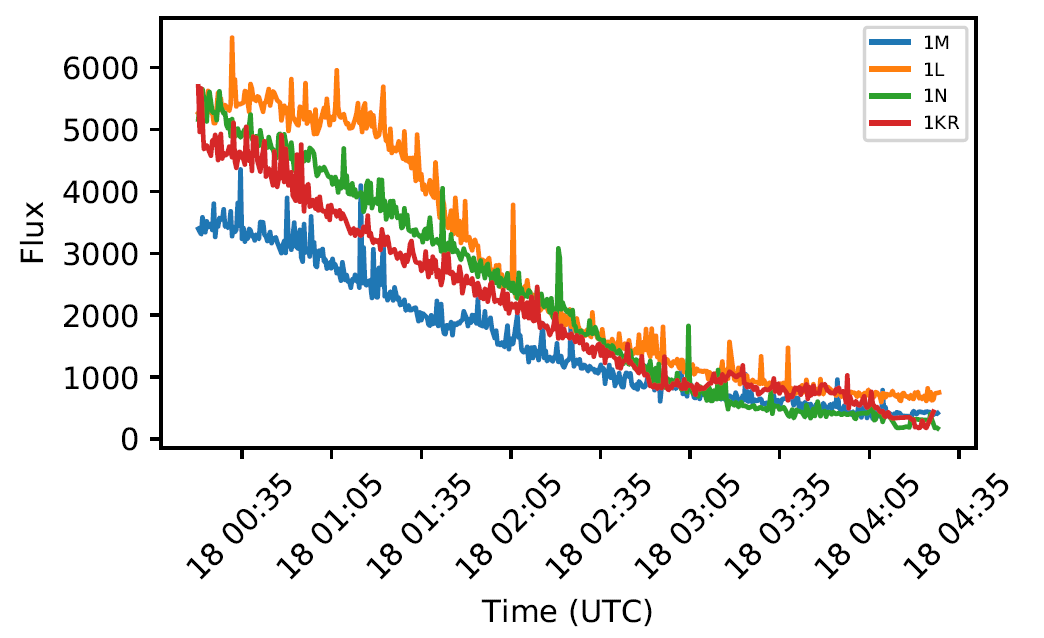}
\vspace{-6mm}
\caption{\small Light curves for Astras satellites (from Winter night observed by SuperWASP) limited to similar period of activity observed from RASA light curves }
\label{fig: fig_9}
\end{figure}

To reproduce the results observed with the RASA dataset, the SuperWASP telescope array observed the Astras cluster over 5 nights, this time for extended periods ($\sim$ 6 hours). The same astrometry and photometry methods were utilised to give the positional plots and light curves shown in Fig \ref{fig: fig_7}. Fig \ref{fig: fig_8} shows the light curve for 1KR over 5 nights and it is evident that the results are reproducible due to the similar signatures of the light curves. Fig \ref{fig: fig_9} shows the light curves isolated to the same period of observation as the RASA light curves and the matching signatures from Fig \ref{fig: fig_6} reaffirm the validity of the unusual features observed. The noise associated with the SuperWASP light curves is significantly greater due to blending, leading to difficulty in isolating small peaks due to some satellite phenomena from noise. This blending is due to the large pixels of the SuperWASP instrument capturing multiple faint stars as well as the Astras satellites. The fluctuations in the background due to these faint stars affect and distort the light curves by corrupting photometry and astrometry[50]. Addressing this error is difficult as it is mostly dependent on the instrument precision, although it could have been partially mitigated by employing more sophisticated source merging prevention measures, such as adjusting the contrast in the deblending procedure during aperture photometry.

\begin{figure}[!t]
\centering
\includegraphics[width=77mm]{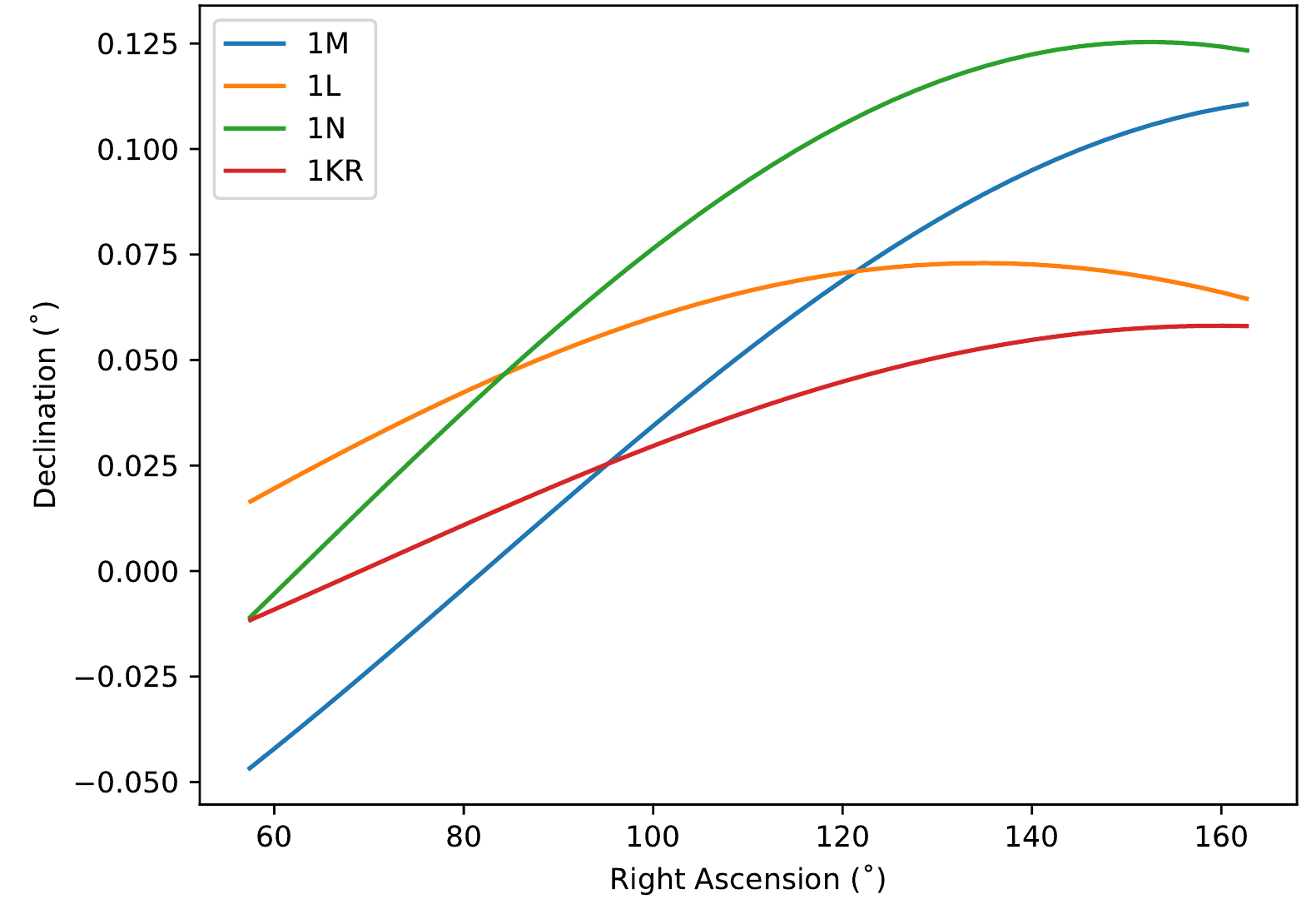}
\vspace{-6mm}
\caption{\small Plot of predicted declination against right ascension for Astras satellites using TLE (two line element) encoded variables for one of the observed Winter nights }
\label{fig: fig_10}
\end{figure}

Since the Astras consist of 2 models of commercial geostationary satellites, namely E3000 by Airbus and A2100 by Lockheed Martin, it is natural to expect the 2 trends observed in the light curves to be associated with the 2 different classes of satellites. Using TLEs near the night of observation, the predicted positional tracks of the 4 satellites were plotted as shown in Fig \ref{fig: fig_10}. Comparing these to the position plots in Fig \ref{fig: fig_5} it is observed that the 2 erratic light curves actually correspond to 2 different models. This dismisses the hypothesis that the contrasting light curve behaviour is due to significant differences in the 2 models and is suggestive that the erratic behaviour is a result of positional and rotational dynamics or some station keeping procedure.

\vspace{-9mm}

\begin{figure*}[!t]
\centering
\includegraphics[width=77mm,height=50mm]{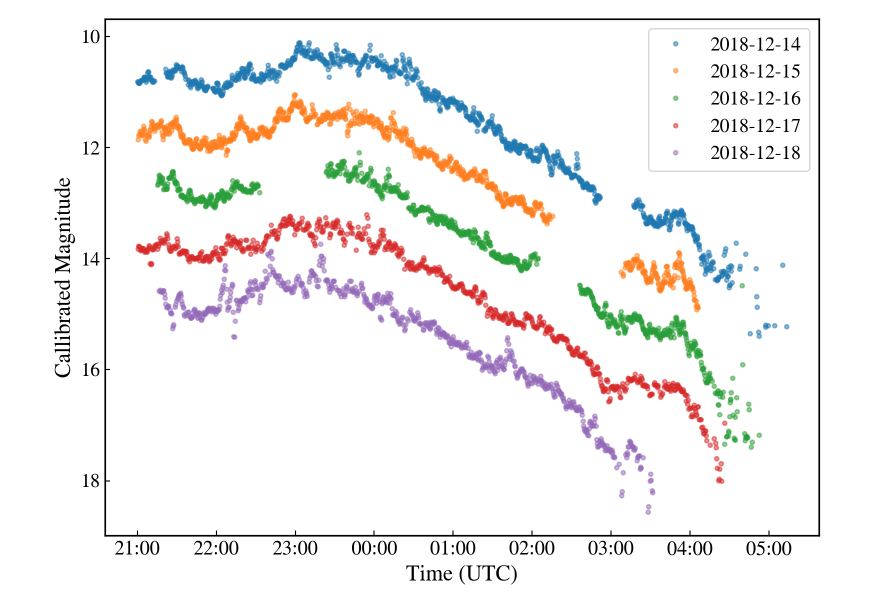} \hspace{3mm}
\includegraphics[width=77mm,height=50mm]{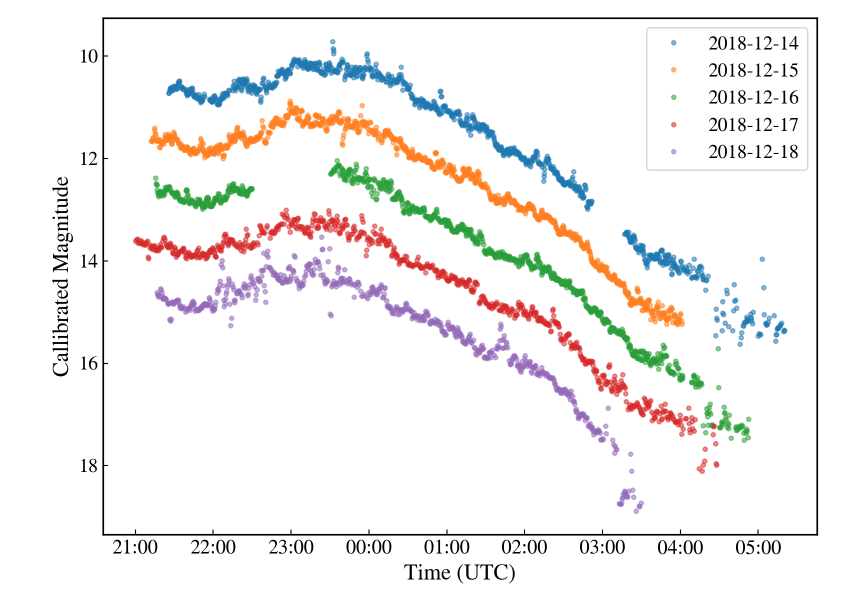}
\vspace{-5mm}
\caption{\small 11a on the left show light curves (calibrated) over 5 consecutive Winter nights for Astra satellite 1L. 11b on the right shows light curves (calibrated) over over 5 consecutive nights for Astra satellite 1N. Observed with SuperWASP instrument in La Palma  }
\label{fig: fig_11}
\end{figure*}

\vspace{2mm}
\subsection{Sun Phase Angle Correction}
The aim of this project shifted from profiling debris to investigating the design and movement parameters of the Astras satellites and so the only parts of the light curves that were of significance were the ones directly caused by the satellite features and behaviour.

Naturally due to the rotation of the Earth and satellites there will be a phase angle (angle between telescope and sun with respect to satellite) variation of the light curves. To remove the effects of this phase angle it was necessary to subtract this natural phase photometric signature from the satellite light curves. However, it is worthy to consider that there is not just 1 fixed signature for a geostationary satellite, as it is variant on the position of the Earth relative to the Sun.

\begin{figure}[!b]
\centering
\includegraphics[width=77mm]{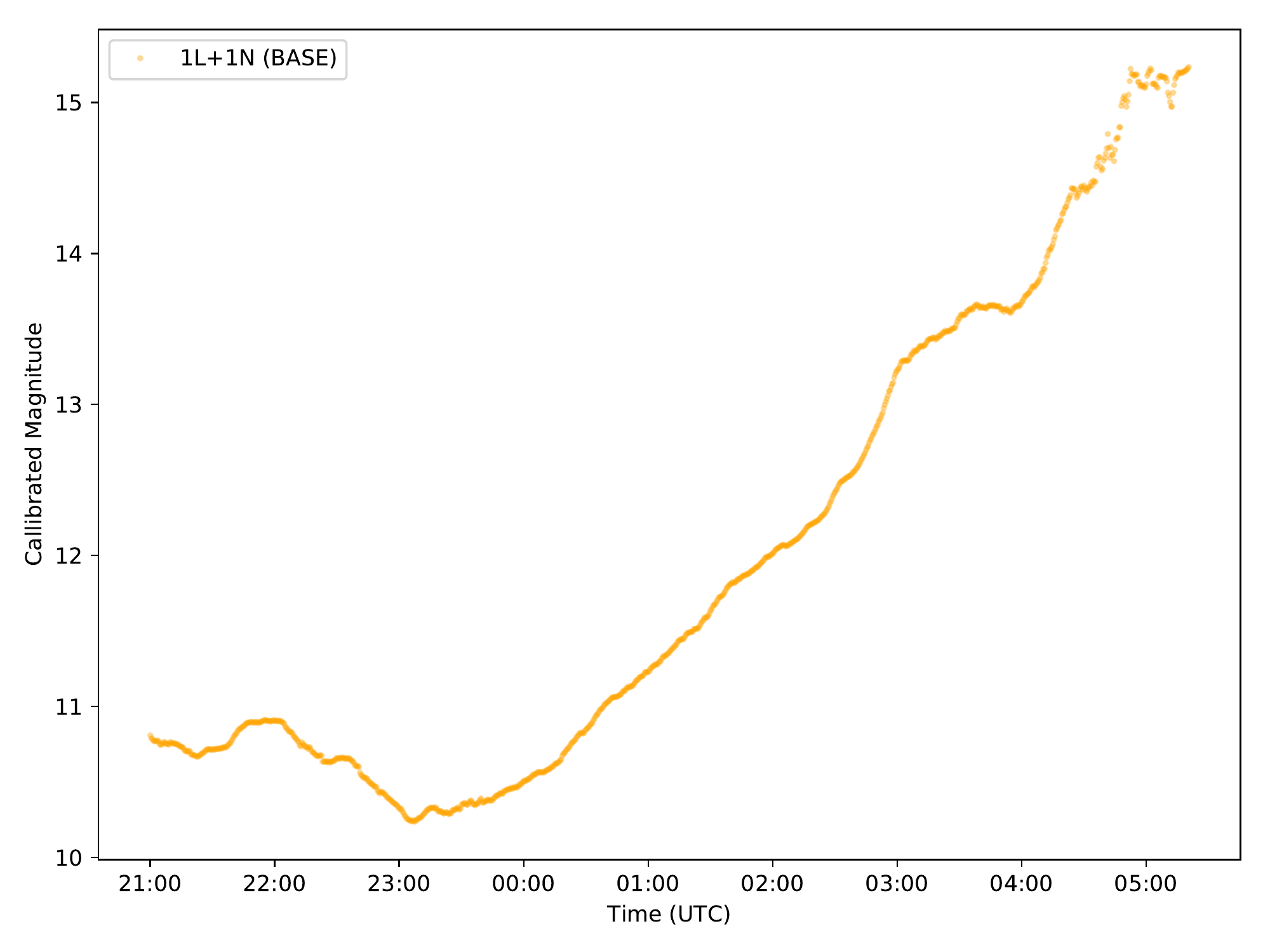}
\vspace{-6mm}
\caption{\small Combined base light curve signature for Winter, composed of aggregate of light curves of Astras satellites: 1L and 1N from 1 night. Non-inverted magnitude scale for comparison with literature  }
\label{fig: fig_12}
\end{figure}

\begin{figure}[!b]
\centering
\includegraphics[width=77mm]{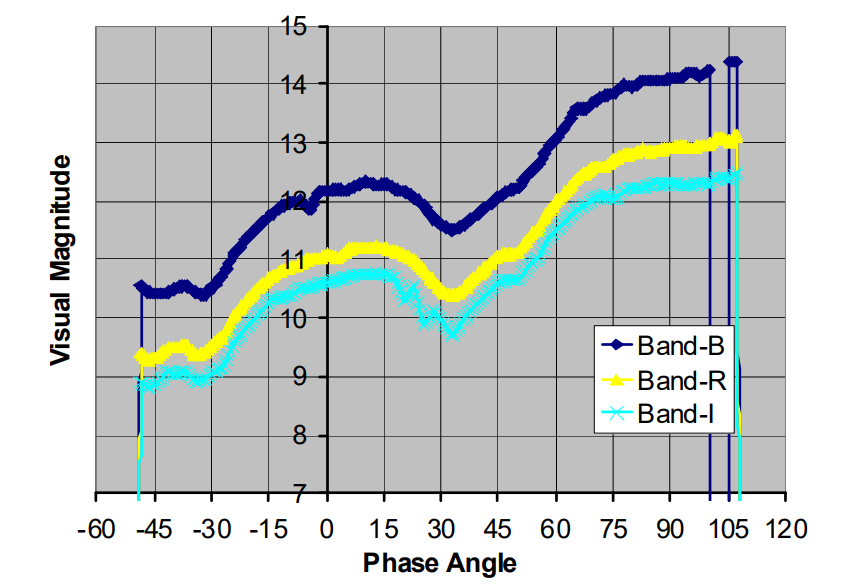}
\vspace{-6mm}
\caption{\small Light curve signatures of AMC 2 during Winter solstice. Observed by the Kirtland Raven telescope. Courtesy of T. Payne and A. Chaudhary [51]}
\label{fig: fig_13}
\end{figure}

\begin{figure*}[!t]
\centering
\includegraphics[width=77mm,height=50mm]{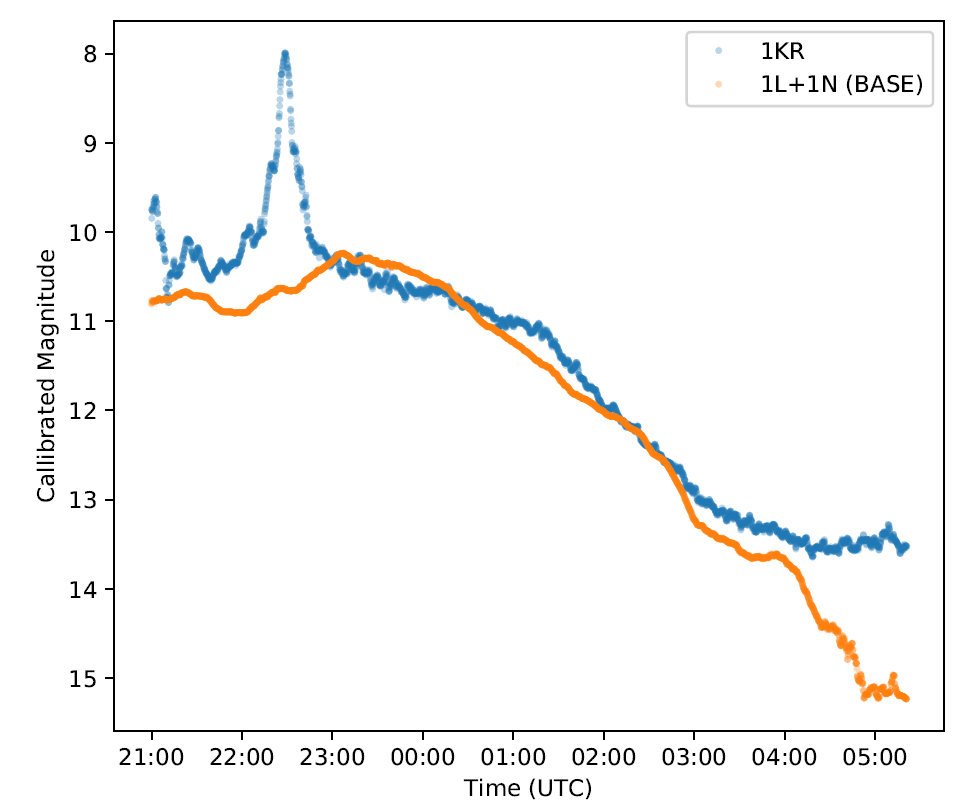} \hspace{3mm}
\includegraphics[width=77mm,height=50mm]{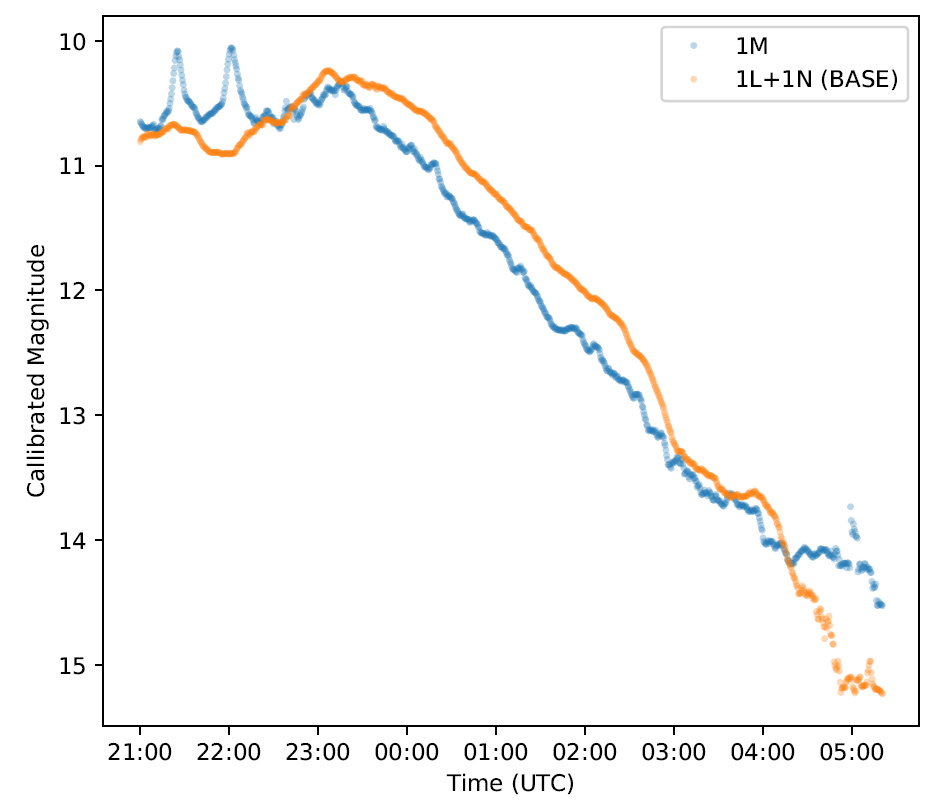}
\vspace{-5mm}
\caption{\small 14a on the left is a plot of the 1L + 1N base signature with the aggregated light curve signature for 1KR (over 5 nights). 14b on the right is a plot of the 1L + 1N base signature with the aggregated light curve signature for 1M (over 5 nights) }
\label{fig: fig_14}
\end{figure*}

\begin{figure*}[!h]
\centering
\includegraphics[width=77mm,height=50mm]{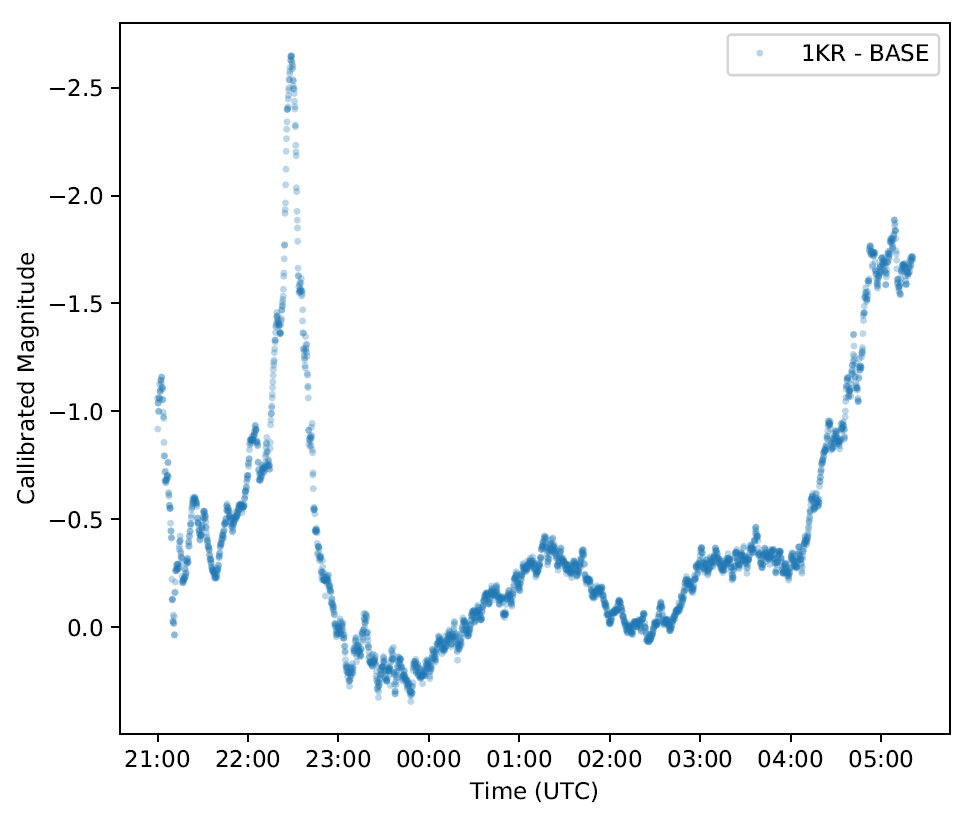} \hspace{3mm}
\includegraphics[width=77mm,height=50mm]{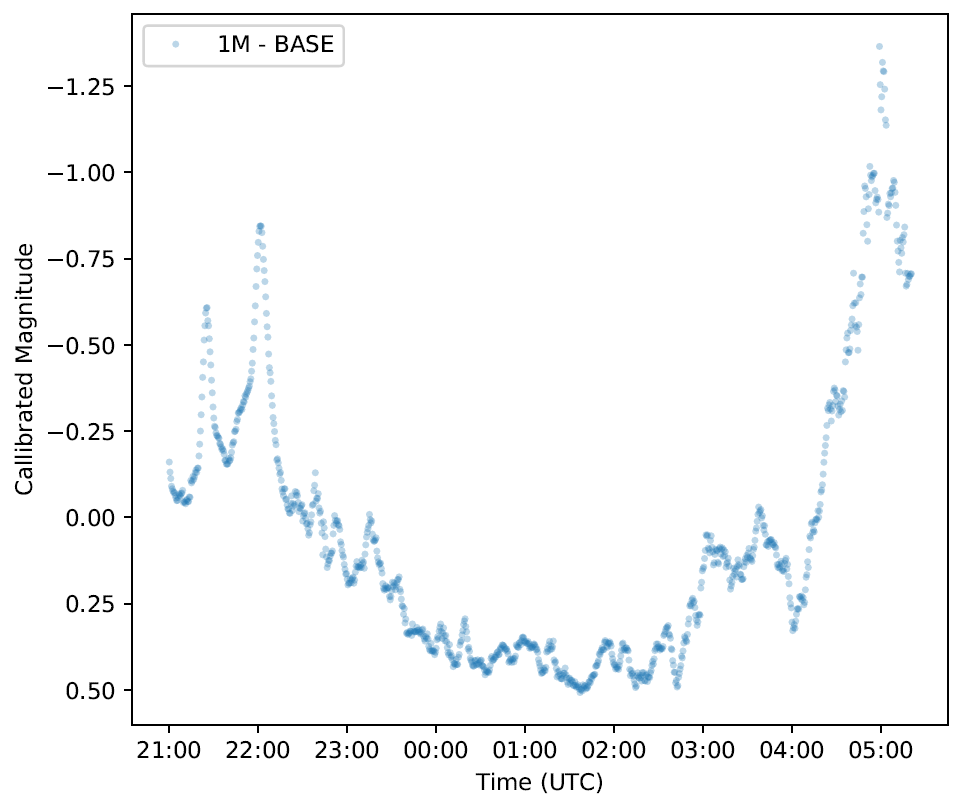}
\vspace{-5mm}
\caption{\small 15a on the left is a plot of the corrected light curve for 1KR, formed by subtracting the base signature from the original 1KR signature. 15b on the right is a plot of the corrected light curve for 1M, formed by subtracting the base signature from the original 1M signature }
\label{fig: fig_15}
\end{figure*}

From Fig \ref{fig: fig_11} it can be observed that satellites 1L and 1N have a similar signature and so their light curves were combined by re-sampling using a rolling mean formulation (to mitigate noise) to construct a base curve as seen in Fig \ref{fig: fig_12}. Note, this relied on a significant assumption that the similar profile of the 2 light curves was indicative of their positional and rotational similarities, despite the fact the satellites are of different models as confirmed from the TLE plots in Fig \ref{fig: fig_10}. This base signature was reaffirmed by Fig \ref{fig: fig_13} from T. Payne and A. Chaudhary's study of American satellite AMC 2 [51] during the winter solstice period: the same period as the SuperWASP observation of the Astras. The base signature is shown against the 2 other satellite light curves in Fig \ref{fig: fig_14} and the overall signatures are similar, reaffirming the choice to combine the 2 light curves for sat 1L and sat 1N. This base profile was subtracted from the light curves of the 2 other satellites to give the corrected light curves seen in Fig \ref{fig: fig_15}. From these it was easier to observe features resulting from only the satellite properties and shape. Fig \ref{fig: fig_15}a is indicative of a repeating peak with a period of $\sim$ 6 hours which corresponds to a phase difference of 90 $\degree$. From the models of the satellites the natural explanation of these peaks are the solar panels becoming fully exposed and then sharply fading out due to their rotation. The 6 hours period then represents the period of rotation of the solar panels.

\begin{figure}[!t]
\centering
\includegraphics[width=77mm]{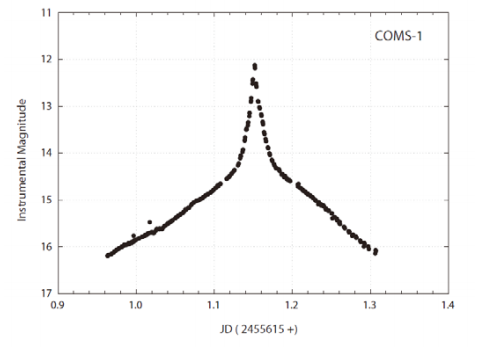}
\vspace{-6mm}
\caption{\small Light curve for Korean COMS-1 satellite from 1 Spring night. Observed from the Ritchy-Chrétien telescope. Courtesy of H. Jin et al. [53]  }
\label{fig: fig_16}
\end{figure}

From Fig \ref{fig: fig_15}b 2 dips can be observed between 21:00 and 22:00 for 1M which are likely results of a shadow being cast on part of the satellite, possibly one of the wide solar wings on the main body or other solar wing. Due to the appearance of 1 major peak it is also possible that the satellite only has 1 solar array, similar to the Planck satellite design[52] despite the reference model for the E3000 featuring 2 wings. Since the reference model images are not of the actual final 1M satellite, this possibility is not to be completely discounted although approached with scepticism due to the unlikelihood of the final satellite being so dissimilar to the reference E3000 model from which it was built. H. Jin et al. also confirmed the single peak light curves of single wing satellites in their 2011 study of COMS-1[53] as seen in Fig \ref{fig: fig_16}. 

\begin{figure}[!b]
\centering
\includegraphics[width=77mm]{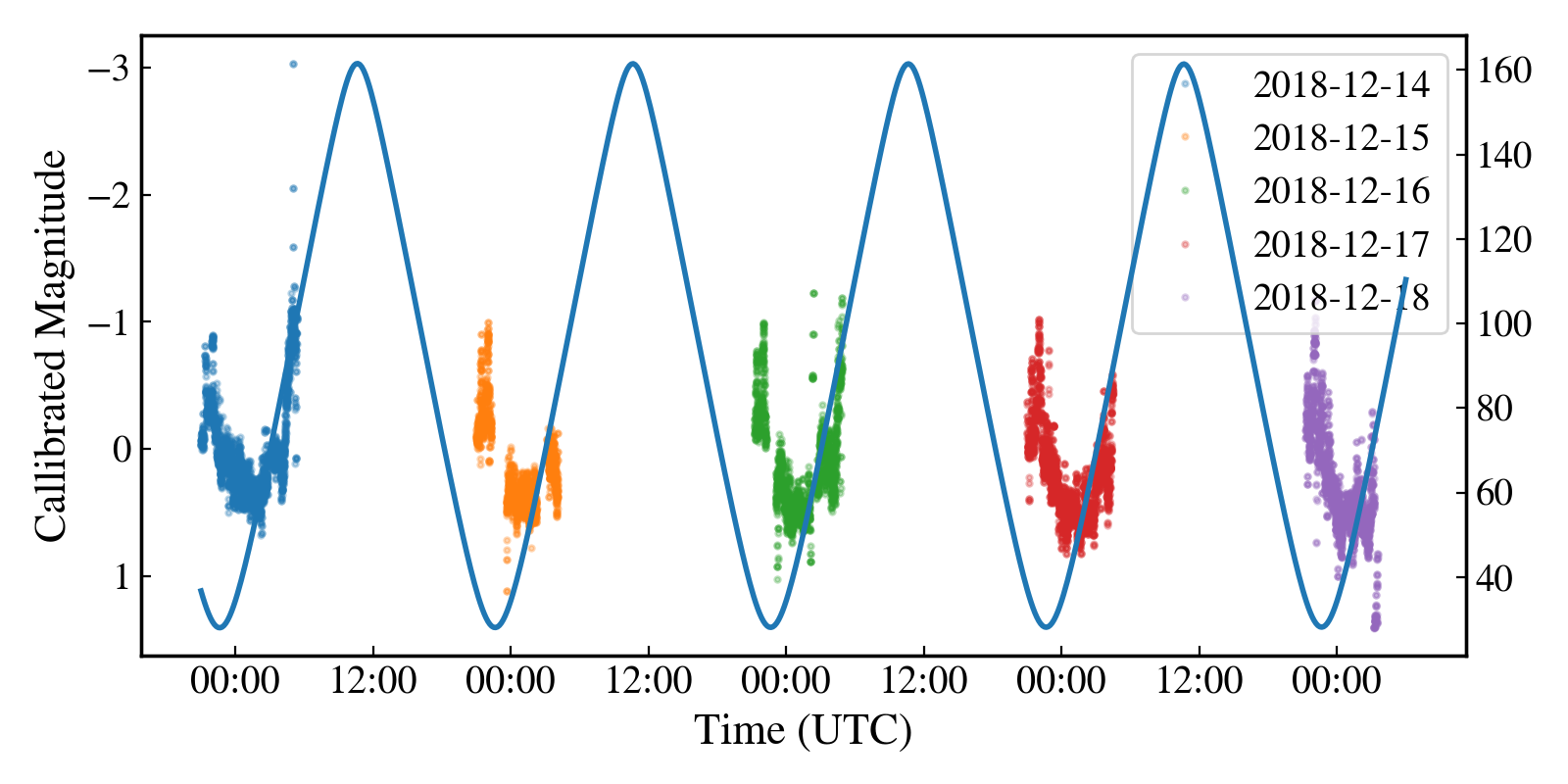}
\vspace{-6mm}
\caption{\small Continuous plot of light curves for satellite 1M over 5 Winter nights with a sinusoidal map displaying periodicity. Observed with SuperWASP instrument in La Palma }
\label{fig: fig_17}
\end{figure}

Due to the limited time of study, it is difficult to fully identify where the subsequent peaks for the 1M satellite occur seen by the cutoff on Fig \ref{fig: fig_15}b. With more time and study resources it would have been beneficial to study for longer than the 8 hour intervals(though difficult with number of night hours), or at different seasonal times. Observing the repeating nature of the light curves over the 5 consecutive nights, it is possible to predict their fully daily periodic behaviour by a sinusoidal map as demonstrated in \ref{fig: fig_17} for satellite 1M using a model 12 hour periodicity. 

\begin{figure}[!b]
\centering
\includegraphics[width=77mm]{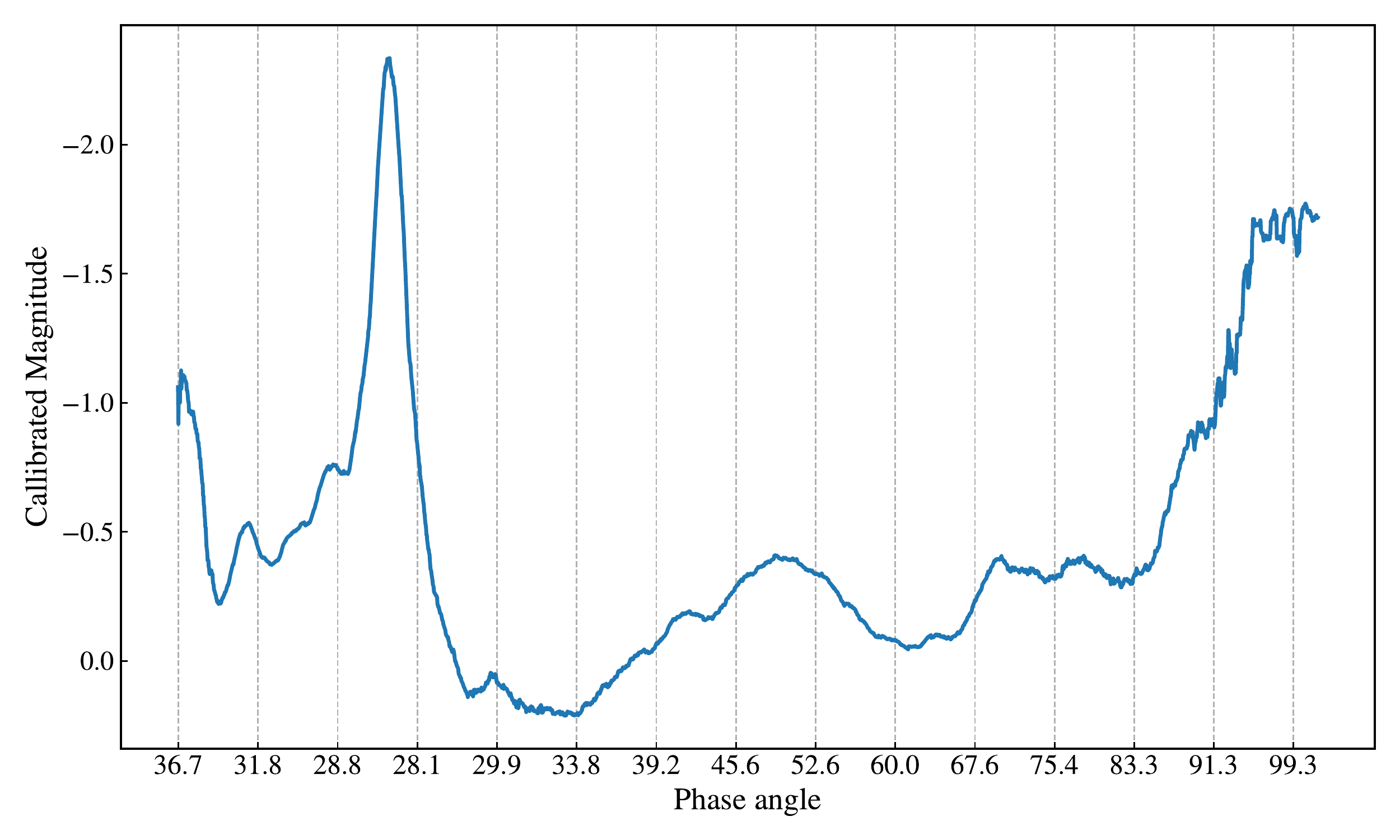}
\vspace{-6mm}
\caption{\small Light curve of satellite 1KR with respect to the sun phase angle. Maximum brightness observed at $\sim$ 28 $\degree$ as predicted }
\label{fig: fig_18}
\end{figure}

\begin{figure}[!t]
\centering
\includegraphics[width=77mm]{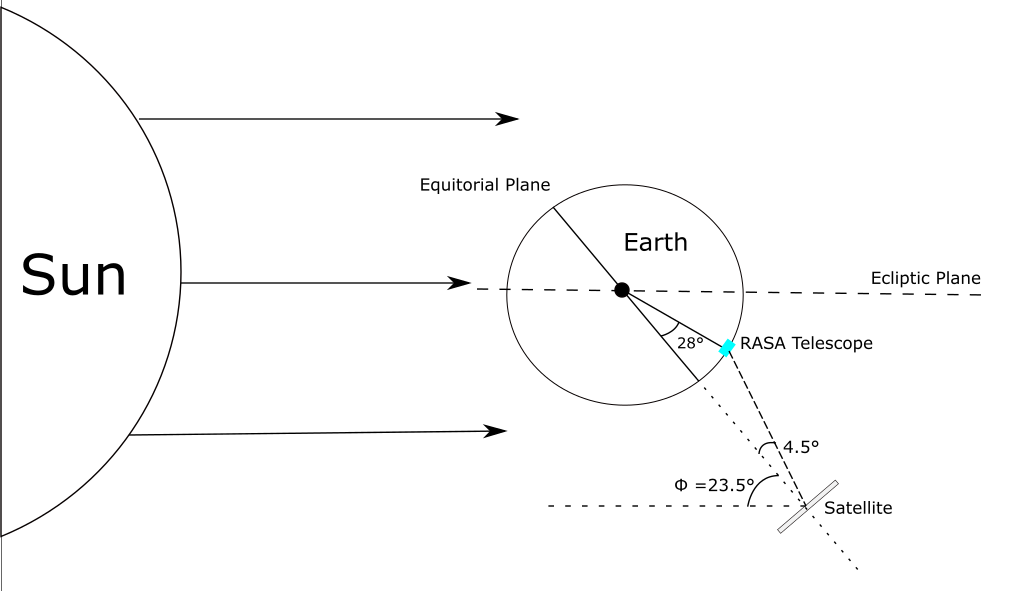}
\vspace{-6mm}
\caption{\small Schematic of Sun-Earth-satellite alignment for an Astra satellite, shown for minimum phase angle as observed by RASA telescope(28$\degree$) }
\label{fig: fig_19}
\end{figure}

From TLE parameters, it was possible to scale the light curves to the sun phase angle instead of time. The minimum phase angle corresponded to a phase angle $\sim$ 28$\degree$ shown by Fig \ref{fig: fig_18}. This is expected due to the 23.5$\degree$ contribution from the axial tilt of the Earth[54] and the other 4.5$\degree$ due to the latitude of the La Palma observatory (28$\degree$N) as demonstrated in Fig \ref{fig: fig_19}. It is also observable from Fig \ref{fig: fig_18} that the minimum phase angle is also representative of the maximum brightness, which is reaffirmed by the studies of R. Buchheim on `Asteroid Phase Curves'[55].

\subsection{Errors, Limitations and Improvements}
This project was hindered in completing its reevaluated goal of explaining the light curves and estimating models of the 4 Astras 19.2$\degree$E satellites, mainly due to the lack of knowledge regarding their dimensions and materials beforehand. Due to the commercial confidentiality preventing public knowledge or even research disclosure, it was difficult to make confident assumptions or utilise more novel methods such as light curve inversion[45], which require albedo models for the objects of interest. Despite this, it was possible to determine and validate light curves for each satellite and then attempt to remove the solar phase angle effect to observe trends and predict explanations.     

\vspace{-5mm}

\vspace{4mm}
The main source of error in this project was the assumption of combining the light curves of Satellite 1L and 1N as beyond observed similarities in their light curves, there was no literature backing up this method. However, in order to assimilate a base phase angle signature relevant to the SuperWASP data, this was necessary and eventually useful in determining the periodicity of sat 1KR's peaks and the potential shadow dips for sat 1M. 

Naturally there was some computational errors, largely due to blending of sources although these were more prevalent for the SuperWASP data over the RASA due to the much greater angular scale per pixel. These could be addressed in future, especially when dealing with debris by employing point spread photometry over aperture, due to its better precision and ability to identify smaller magnitude objects (debris). The most significant measure to mitigate this blending impact would be to have better instruments for observation, though this is financially constrained. The RASA is suited for this, however, it compromises a larger frame of view and file space due to more detailed images.

\subsection{Further Research}
The overarching aim of this project was to develop a basis to study debris precisely in the geostationary orbit by firstly studying the objects of significance: satellites, and developing computational methodology that can then be transferred to debris data with appropriate modifications. Due to limitations, the objective changed to explaining the behaviour of the Astra 19.2$\degree$E cluster from their photometric light curves. However, this project gave important foundations for work towards the real problem of space debris.

As discussed before the difficulty in tracking geostationary debris is not even its quantity, but the size and velocity of it making precise measurements difficult. Currently radar mechanisms such as the Tracking and Imaging Radar (TIRA)[56] employed by the European Space Agency (ESA) are able to detect space debris, however their degree of accuracy is incredibly limited. They utilise methods such as ``beam parking"[57], focusing on a region of sky for many hours with objects crossing the frame of view measured for ``direction, range, range rate" from which information about their size and orbit can be estimated. The accuracy of this as mentioned before is extremely limited; unsuitable for addressing the risk of debris-satellite collisions. 

The focus for geostationary debris profiling is shifting to laser tracking of space debris due to the `inherent accuracy'[58] associated with laser. This is a significant advantage over the predecessor radar methods which lack the accuracy needed. These methods were tested at the Stromlo SLR[58] with success in regards to quick and accurate orbit determination, however ``optimisation of the real-time orbit quality and down-range tracking network configuration" requires further research.

\section{Conclusion}
The GEO belt's uniqueness has unsurprisingly lead to it being labelled ``space's most valuable real estate"[59]. The ability to stay fixed over a certain point means tracking and receiving antennas on Earth do not need to track satellites in this orbit, making them an invaluable asset to commercial and military communications. Considering their reliance as well as their construction and enabling costs, it is evident that their operational safety is of paramount importance. 

This project has gave insight into the Astras 19.2$\degree$ cluster and developed methods for studying their behaviour through photometry. These methods have the capacity to be utilised for debris study provided appropriate data, however their effectiveness is unlikely due to necessity for fast, instantaneous determination of size and orbit for debris. Instead, these developed methods are more useful for studying other geostationary satellites and to identify the effect of debris collisions from unexpected deviations in the satellite light curves.

The dilemma of debris is better left to more precise and practical tools such as the aforementioned laser tracking. The goal of achieving the extreme precision needed for safely profiling debris can only be achieved with such revolutionary methods, although they are not without problems of their own. As discussed, the main problem with GEO debris is not the quantity of it, but rather with how poor the precision of detecting it is; so photometric study and even radar are simply incompatible.

What remains inevitable is that the geostationary satellite population will only increase (EUMETSAT[60]) further, especially as technology wanes towards the `5th Generation'. But what is necessary is an equivalent growth in debris profiling and management research, as well as a more serious commitment to suitable satellite decommissioning. Without this, the danger of debris on these essential satellites will only exacerbate, with more violent collisions leading to even more expelled debris. Whether it be by laser tracking through LEO instruments or novel technology that is not substantiated yet, the problem of profiling geostationary debris must be addressed treated with imminent importance.

 
\end{document}